\documentclass[prd,aps]{revtex4}
\usepackage{psfig}
\newcommand{\beq}{\begin{equation}}
  \newcommand{\eeq}{\end{equation}}
\newcommand{\beqa}{\begin{eqnarray}}
  \newcommand{\eeqa}{\end{eqnarray}}
\newcommand{\beqar}{\begin{eqnarray*}}
  \newcommand{\eeqar}{\end{eqnarray*}}

\def \m {{\bf m}}

\def \la {\langle}
\def \ra {\rangle}

\def \s {\,\,\,\,}

\def \e {e}
\def \Eo {E_o}
\def \m {m}
\def \m1 {m_1}

\def \Er {E_{\cal R}}
\def \Es {E_{\cal S}}
\def \Sr {S_{\cal R}}
\def \Ss {S_{\cal S}}
\def \R {${\cal R}$}
\def \S {${\cal S}$}

\def \SS {${\cal S}$ }

\begin{document}
\title{Thermodynamics with long-range interactions: 
  from Ising models to black-holes}

\author{Jonathan Oppenheim$^{(1)(2)}$}

\affiliation{$^{(1)}$
  Racah Institute of Theoretical Physics, Hebrew University of Jerusalem, Givat Ram, Jerusalem 91904, Israel}
\affiliation{$^{(2)}$Theoretical Physics Institute, University of Alberta, 412 Avadh Bhatia Physics Laboratory, Alta., Canada, T6G 2J1.
}
\begin{abstract}
  New methods are presented which enables one to analyze the thermodynamics of
  systems with long-range interactions.  Generically, such systems have entropies which 
  are non-extensive,
  (do not
  scale with the size of the system).  We show how to calculate the degree of non-extensivity for 
  such a system.  We find that a system interacting with a heat
  reservoir is in a probability distribution of canonical ensembles.
  The system still possesses a parameter akin to a  
  global temperature, which is constant throughout the substance. 
  There is also a useful quantity which acts like a {\it local temperatures} and
  it varies throughout the substance.  These quantities are closely related
  to counterparts found in general relativity.
  A lattice model with long-range spin-spin coupling is studied.  
  This is compared with systems such as those
  encountered in general relativity, and gravitating systems with Newtonian-type interactions.
  A long-range lattice model is presented which can be seen as a black-hole analog.  
  One finds that the analog's temperature and entropy have many properties
  which are found in black-holes. 
  Finally, the entropy scaling behavior of a gravitating perfect fluid of 
  constant density is calculated.
  For weak interactions, the entropy scales like the volume of the system.  As the 
  interactions become stronger,
  the entropy becomes higher near the surface of the system, and becomes
  more area-scaling. 

\end{abstract}
\maketitle

\section{Introduction}
In the study of thermodynamics, it is almost always implicitly assumed that the system
does not possess long-range interactions.  Very little is known about the thermodynamics
of systems which do possess long-range interactions, except in special cases such as
plasmas where the electromagnetic interactions are screened, or systems which have
no overall charge\cite{landlieb}.  In both these instances, one can use standard thermodynamics,
since effectively, there is no long-range interaction.  
If however, the long-range interactions are not screened, then difficulties are encountered,
such as the non-existence of the canonical ensemble\cite{paddy} or
inequivalence of microcanonical and canonical ensembles, and
potential lack of a stable equilibrium configuration\cite{antanov,lynden1}.
The latter is sometimes attributed to negative heat-capacities\cite{thirring}.  
Negative heat capacities are not only present in astrophysical systems,
but have even been observed in fragmenting nuclei\cite{negcap-nuc} and
atomic clusters\cite{negcap-sod}.
It is not known how to deal with these systems
generically, although there have been some attempts to understand
them outside of standard thermodynamics using the Tsallis
entropy\cite{tsallis} [c.f. also \cite{renyi}].

A principle
motivation for this work is therefore to provide a framework
in which to study such systems.  A second motivation comes
from the study of black-hole thermodynamics.  There, it is
found that the black-hole posseses an entropy which has unusual
properties.  Here, we will show that these properties are not
limited to the black-hole, but that other systems with long-range
interactions exhibit related behavior.  We will essentially construct
an analog of a black-hole by adding long-range interactions to a spin-lattice
model.

Systems with long-range interactions are often referred to as {\it non-extensive}, because
the entropy and energy do not scale with the volume of the system.  
Normally, if one has a thermodynamical system, and one holds the intensive
variables (temperature, pressure, and chemical potential) fixed,
then if the size of the system is doubled, the extensive variables (entropy
and energy) 
will also
double.  This is not true if the interactions are long-range.

The purpose of this article
is to develop new methods and a formalism to explore a number of facets of such non-extensive systems.  
We shall
employ a principle of "local-extensivity" which enables one to define 
thermodynamical quantities 
for an interacting
system.  We shall also show how to classify the degree of non-extensivity of the system, by
calculating the scaling behavior of the entropy as a function of total energy.
The motivation for this part of the study comes from general relativity.
There, it is found that the entropy of a black-hole is proportional to its
area, rather than its volume (i.e. the entropy is non-extensive).  In this
study, we will see that this is a generic property of interacting systems,
rather that something unique to the black-hole.

We will also see that generically, a system interacting with a reservoir
is not at a particular temperature, but rather, is in a probability
distribution of temperatures.
This will be found by studying a system interacting with a reservoir in the
microcanonical ensemble i.e. the total energy of the system plus reservoir
is fixed. 
Usually, if one then only looks at
the system, it will be in a canonical ensemble (fixed temperature).
When interactions are present, this will not be the case.
This leads us to introduce a new type of ensemble,
which we call the microlocal ensemble.  It is equivalent to the microcanonical
ensemble when there are no interactions.

Despite the fact that a system is not found in the canonical ensemble,
we shall see that one can define a quantity we call the {\it global
  temperature} $\beta_o$.  It describes the system as a whole,
and is written in terms of the total energy, including the interacting terms.
There is also a {\it local temperatures}, a quantity inspired by
general relativity.  Both types of temperature are measurable in principle.

In the case of short-range interactions, two systems brought into thermal contact will be at the same
temperature.  
Here, we will show that also for interacting systems, the global temperature 
of two systems
is the same. This justifies to some extent, our use of the term
temperature to describe $\beta_o$.  However, we will see that if one has
two separated systems with the same global temperature, then when they
are brought into contact adiabatically, they will reach a new global
temperature.  $\beta_o$ is therefore not an intensive quantity.  We will also see that
the local temperatures of the two systems are in general different for each
system when in thermal equilibrium.  
Such an effect is analogous to the
Tolman relation\cite{tolman} which exists in curved space.  
There, one finds a temperature gradient
due to the curvature of space-time.
Essentially, frequencies are red-shifted by curvature.  Since the temperature gives the probability
distribution of a frequency spectrum, it is also red-shifted.
Here, we see that such an effect does not exist solely in curved space-time,
but can also be
thought of as due to the presence of long-range interactions.

To make our discussion more concrete, we will examine a toy model
consisting of a lattice of spins in a magnetic field,
and interacting via a spin-spin coupling.  However, rather than only nearest neighbor interactions,
we will also consider the long-range couplings.  We will consider the case
of a uniform long-range interaction, as well as the case of two different
systems interacting via two unequal uniform interactions.  Such a situation
arises when one considers two lattice clusters which are of small size.
We will also discuss the continuum situation, where the interaction can be
arbitrary, and varies from site to site.

We then consider thermodynamics in
the general theory of relativity.  Comparisons between the lattice model and of black-hole thermodynamics provide
another motivation for this study.  Previously, analogs of
black-holes\cite{dumbhole}\cite{visser},
(so called, acoustic, or solid state black-holes) have been used to understand black-hole radiation.
However, they are not useful for understanding the black-hole entropy.
Here, we see that one can construct a a black-hole analog that can be used to study black-hole entropy.
One 
finds that the entropy can be non-extensive, just like a real black-hole.
We find for the analog that there is an infinite red-shifting between its local temperature
and its global temperature which has exactly the same form as a black-hole.  
At exactly the point where the systems acts like a
black-hole, a
degeneracy in the local energy levels forms.  This degeneracy is universal,
in the sense that it only depends on the form of the interaction. 
The
universality is somewhat reminiscent of the universality of black-hole
entropy.  

We will also investigate other gravitational systems in general relativity.  
In particular, we look at the entropy
scaling behavior of a gravitating perfect fluid.  The motivation for this comes partly
from an earlier study\cite{area} where I showed that the black-hole
is not the only gravitating system which has an area-scaling entropy.
A system of shells has an entropy that
scales as the volume when the gravitational interaction is weak, but
the entropy becomes area scaling at the point before a black-hole is
formed.  Here, in looking at the gravitating perfect fluid, we find related behavior.
We can look at the entropy scaling behavior not just in limiting cases,
but for all strengths of the gravitational interaction. 
As the strength of the gravitational interaction is increased, the entropy
slowly moves to the outer surface of the perfect fluid 
One 
finds that the total entropy is non-extensive, just like in a black-hole, and
approaches area scaling behavior as the strength of the gravitational
interaction gets stronger.  
 
We also explore gravitating systems in the context of Newtonian-type dynamics.
This is done to show that
the red-shifting of temperatures -- usually considered to be 
an effect related to the curvature of geometry --  also exists in other
gravitational models which are not geometric theories. 

The paper is organized as follows.  In Section \ref{sec:longrange}
we introduce our formalism.  First, in Subsection 
\ref{ss:localextensivity} we introduce our assumptions,
which we call {\it locality} and {\it local extensivity} and show that these assumptions
are obeyed by a number of common systems.  Next, in Subsection
\ref{ss:temp} we use the microcanonical ensemble to show that a system
interacting with a reservoir will not be found at a particular temperature,
but rather, will be in a probability distribution of different temperatures.
Nonetheless, there is a parameter which behaves very similarly to
a temperature, which we call the global temperature.  This is defined
in Subsection \ref{ss:global}.  The physical significance of the global
temperature, as well as another parameter called the local temperature
is explored in Subsections \ref{ss:physsigofglobal} and
\ref{ss:physsigoflocal}.  Then, in Subsection \ref{ss:unequal}
we show that the global temperatures of two systems brought into
contact, are equal at equilibrium.  The local temperatures need not
be equal (an effect reminiscent of red-shifting which is usually 
considered to be the sole domain of general relativity).  This allows us
to study lattice models where the long-range interaction is not uniform.
Next, in
Section \ref{sec:exising} we explore in some detail a lattice
model with long range interactions.  In Section \ref{sec:bhanalog}
we show that such a system can be made into an analog of a black-hole
and has a temperature and entropy with  many properties reminiscent of black-holes.  
In Section
\ref{sec:scaling} we show how one can generically calculate the entropy
scaling behavior of an interacting system.  This is done for a gravitating
perfect fluid in Section \ref{sec:entropyfluid}.  We find that the entropy
becomes more area scaling as the gravitational interaction gets stronger.   We conclude
with some general remarks in \ref{sec:conclusion} and point to
some open questions.

  In Appendix
\ref{ap:newtonian} we look at systems under the influence of Newtonian-type
gravity, and show that an analog of the Tolman relation exists -- local
temperatures are red-shifted.
In Appendix \ref{ap:continuum} we look at more general interactions
and go to the continuum limit.

%
\section{A formalism for systems with long-range interactions}
\label{sec:longrange}
%
\subsection{Locality and local extensivity}
\label{ss:localextensivity}
%
Let us consider two interacting systems, $1$ and $2$
with total energy
\beq
m=E_1+E_2 +G(E_1,E_2)
\label{eq:int}\s.
\eeq
Here $G(E_1,E_2)$ is some interaction potential (which may
include self-interacting terms) and $m$
is the total energy\cite{notation}.  
%
In the absence of the interaction $G$, the energy of each
system would be $E_1$ and $E_2$.  We will refer to $E_1$
and $E_2$ as the
{\it local energy}.  Likewise $E=E_1+E_2$ is the local
energy of both systems.  In other words, $E$ can be thought
of as the extensive part of the energy (or non-interacting part).

In general, a complicated system
will include many such systems interacting together, or will be a 
continuum of separate systems at each point in space.  
For the purposes of illustration, we will consider
the simple case where we can divide the total system into two parts.  We will later consider 
more complicated set-ups.

We now make two assumptions.\\
{\bf Locality:} {\it the state of each system is determined only by its local
  energy $E_i$ and local variables $q_i$.} \\
{\bf Local extensivity:} {\it correlations due to the interaction can be completely 
  described by correlations in the energies $E_i$.}

The first assumption is rather generic.  The latter assumption is true for two systems interacting
via a potential which can be put into the form Equation (\ref{eq:int}).  
The essential requirement for local extensivity is that if the
potential introduces correlations in microscopic variables $q_i$, 
then it will also result in correlations
between the local energies $E_i$.  
Correlations which depend on other thermodynamical quantities can also be
described using this formalism.  The logic behind these assumptions should become 
clear in a moment when
we consider an example, but an important consequence is that
if we write the total entropy of the two systems as $S(E_1,E_2)$ 
then
\beq
S(E_1,E_2)= S_1(E_1)+S_2(E_2)
\label{eq:additive}
\eeq
where $S_i$ is the entropy of each subsystem.
This relation
is somewhat counterintuitive, because when interactions are present one expects there to be
correlations between the two systems, and therefore, one does not expect the entropy to be additive.
However, the entropy is only additive because we have written it in terms of the local energies.
The entropy does not scale linearly
with the total energy $m$, and is therefore non-extensive.
Essentially, for fixed $m$ there are correlations
which exist because $E_1$ is not independent of $E_2$, but once you 
specify $E_1$ and $E_2$ you have completely
specified each subsystem.

Equation (\ref{eq:additive}) follows from our assumptions, because if System
1 and 2 are only determined by local variables, then specifying local
variables, such as $E_1$ and $E_2$ determines the number of possible states
of each system.  Furthermore, since the correlations between the two systems
are only correlations between values of $E_1$ and $E_2$, then once $E_1$
and $E_2$ are specified there are no additional correlations which
would destroy the additivity of the entropy as given by Eq.
(\ref{eq:additive}).

To make this point clear,  let us illustrate it with an example: consider  
a lattice of $N$ spins with total energy
\beq
m=\sum_j h_j \sigma_j - \sum_{\la jk\ra} J_{jk}\sigma_j\sigma_k
\label{eq:ising}
\eeq
where the $\sigma_j$ represent the spin at each lattice site (with values $\pm 1$), and the 
$h_j$ are magnetic field values(or internal energy levels).
The $J_{jk}$ are  spin-spin coupling constants.  In the standard Ising model, one takes the sum such
that $\la jk \ra$ are pairs of nearest neighbors.  Here, we consider the case where the spin-spin coupling
is strong enough (or the lattice spacing small enough), that $J_{jk}$ and $h_j$ are relatively constant over a large region.
For simplicity, we will imagine that the system is
composed of two such regions separated by a short distance.  This then gives (up to a constant)
\beq
m=h_1 e_1 + h_2 e_2 -J_1e_1^2/2 -J_2 e_2^2/2 -J_{12}e_1e_2
\label{eq:twoissings}
\eeq
where the dynamic variable $e_i$ is the number of up spins minus the number
of down spins inside each region, and
$J_i$ and $h_i$
are the coupling inside each region and are known constants.  
$J_{12}$ is the coupling between each region, and would presumably be
smaller than the $J_i$.
The number of sites in each region is assumed constant.

Now it is clear that our assumptions, and Equation (\ref{eq:additive}) hold.  
Since the local energy of each system is $E_i=h_i e_i$,
specifying $E_i$, completely fixes the number of up and down spins in each region.  
Furthermore, once $E_1$ is specified,
then the state of System 1 is completely independent of the second system.  
I.e., once $E_1$ is specified, the state of System 1 has been determined
(macroscopically).  This state will now not depend on what  
value of $E_2$ System 2 happens to have.  Different values of $E_2$ will
of course mean that the effective magnetic field that System 1 feels will
be different, but we have already specified $E_1$, so its macroscopic state
will not change.  Its microscopic state won't change either, since the
interaction of Equation (\ref{eq:twoissings}) doesn't introduce any
distinction between different microscopic configurations.  
Specifying $E_i$ is the same as specifying $e_i$ since the $h_i$ are known.
For each system, once the spin-excess $e_i$ is specified, then all spin
combinations consistent with this value of $e_i$ are equally likely.  Once
again, once $e_1$ is specified for System 1, nothing depends on what happens
with System 2 insofar as which states will be occupied. 

The entropy of each system $S_i$ is just given by the number of
independent ways of arranging the spins (since each arrangement is equiprobable).  I.e.
\beq
S_i(E_i)= - \frac{N_i+e_i}{2}\log\frac{N_i+e_i}{2N_i}
- \frac{N_i-e_i}{2}\log\frac{N_i-e_i}{2N_i}
\label{eq:spinentropy}
\eeq
There are of course, still correlations between the two systems -- for
fixed $m$, the number of up spins in System 2 will
completely determine the number of up spins in System 1.

We will see that these assumptions also hold in the context of general relativity, but for the moment, let us
return to the generic case and define the temperature.  One has to be careful, because as we will see, the
canonical ensemble does not exist.  

\subsection{Multiple temperatures and the microlocal ensemble}
\label{ss:temp}

The usual derivation of the canonical ensemble follows from considering a
large  
reservoir \R\ in contact with a 
smaller system \S.  One fixes
the total energy of the combined system 
(hence one is operating in the microcanonical ensemble), but we
let energy flow between \R\ and \S.  
One then finds that the probability distribution of energies of 
\S\ is independent of the
details of the reservoir.  The
distribution depends on a
quantity which is defined as the temperature and this
defines the canonical ensemble.
Here, we essentially repeat the standard derivation, except  we
have the interaction term given in Equation (\ref{eq:int}).  We will find
that the distribution looks very different.  One can think of the system
as being in a probability distribution of different canonical ensembles.

Although we could work in the microcanonical distribution from the very
start, it will prove useful to define a new ensemble which we will call
the {\it microlocal ensemble}.  Rather than fixing the total energy $m$,
we shall fix the total local energy $E=\Er+\Es$.  The motivation for this should
become clear as we proceed.  In the case when the interaction $G$ is zero,
the microlocal ensemble and the microcanonical ensemble are clearly
identical.

Let us therefore consider two system with fixed local energy $E$,
and imagine that \R\ is very large, and constitutes a reservoir i.e. $\Es
\ll \Er$.
We then allow energy to flow between \S\ and \R\ until the systems reach 
equilibrium.  At equilibrium, and for large systems, one is most likely
to find the system in a state which maximizes the entropy.
   
The probability that \S\ has energy $\Es$ for a fixed $E$ is given by 
counting the number of possible states of the system when
\SS has local energy $\Es$ and \R\ has local energy $\Er=E-\Es$.  
The probability that \S\ has energy $\Es$ for a fixed $E$ is a
conditional probability and is denoted by $p(\Es|E)$.  I.e. it is the
probability of having energy $\Es$ conditional on the total energy being $E$.
We can write
\beqa
p(\Es|E)d\Es&=&\frac{\Omega_S(\Es)\Omega_R(\Er)d\Es}{Z_E}\nonumber\\
&=&\Omega_S(\Es) e^{\Sr(\Er)}d\Er/Z_E
\eeqa
where $\Omega_S(\Es)$ and  $\Omega_R(\Er)$ are the number of states of \S\
and \R\ with energy $\Es$ and $\Er$.
$Z_E$ is the partition function obtained by counting all states with a
fixed $E$
\beq
Z_E=\int_E d\Es \Omega_S(\Es)e^{\Sr(\Er)}\s .
\eeq
We can now expand $\Sr(\Er)$ around $E$ to give $\Sr(E)-\Es\partial
\Sr(E)/\partial E$. We then define
the inverse temperature $\beta_E$ in the usual manner
in terms of the local extensive entropy
\beq
\beta_E\equiv\frac{\partial \Sr(\Er)}{\partial \Er} \s
\label{eq:localt}
\eeq
We shall refer to $\beta_E$ as the {\it local temperature}.
The motivation for using this term (as with many of the terms we are 
introducing) comes from general relativity.  

Note that the temperature
of the system is defined in terms of the derivative of the
{\it reservoir's entropy}.  In the non-interacting case, no issues
arise from this definition: if two systems are in thermal contact
in the microcanonical ensemble, then 
$\frac{\partial \Ss(\Es)}{\partial \Es} \simeq \frac{\partial \Sr(E)}{\partial
E} $.  When long-range interactions are present, this is not 
necessarily true -- a point which will be discussed in 
Subsection \ref{ss:unequal}.  One therefore should keep in mind
that the temperature is a property of a reservoir -- it gives
the distribution associated with a smaller system in contact
with it. 
In the case where the division of a {\it single} system into
a reservoir and smaller system
is purely formal, we will see that
$\frac{\partial \Ss(\Es)}{\partial \Es} \simeq \frac{\partial \Sr(E)}{\partial
E} $.  This comes from symmetry considerations.
Another special case is when the reservoir has no long-range
interactions.  Both these case will be discussed in Appendix \ref{ap:equalt}.
In general, one can relate $\beta_E$ to $\frac{\partial \Ss(\Es)}{\partial
\Es}$ using the methods we will develop in \ref{ss:unequal}.

Using the definition of Eq. (\ref{eq:localt}), we find
\beq
p(\Es|E)d\Es=\frac{\Omega_S(\Es)e^{\Sr(E)-\Es\beta_E}d\Es}{Z_E}\s.
\label{eq:canon}
\eeq
One can see from the above expression, that at fixed $E$, 
the probability distribution of \S\ looks independent
of the nature of the reservoir, since the probability
for the system to be at a certain energy $\Es$ is
\beq
p(\Es|E)d\Es\propto \Omega_S(\Es)e^{-\Es\beta_E}d\Es \s.
\label{eq:canonprop}
\eeq
which looks like the ordinary canonical ensemble.  We must keep
in mind however, that the temperature is defined in terms of 
the reservoir, not the system.

Now however, there is a key difference. We have worked in the microlocal
ensemble, but in the
microcanonical ensemble, it is not $E$ that is held fixed, but rather $m$.  
If there were no interactions,
then this would mean that $E$ is also fixed (if $G=0$ then in fact, $m=E$).
Then, since $E$ would be constant, so would $\Er(E)$.  We would therefore
recover the fact that \S\ is in the usual canonical ensemble.
However, for $G(\Es,\Er)\neq 0$, $E$ need not be constant.  For example, if
$G(\Es,\Er)$ were
quadratic in $E$, then at fixed total energy $m$ the system can be in one 
of two values of $E$.  For more complicated potentials, 
{\it many values of the local energy $E$ are possible
even though both
$m$ and $\Es$ are fixed}.  Therefore, for an isolated
system, one should not hold $E$ fixed, but rather $m$ as this is the
conserved
quantity, while the local energy $E$ of an isolated system can change.

In order to calculate the probability distribution of the system in the microcanonical
ensemble we can use {\it the law of total probability}
\beq
p(y)=\sum_x p(y|x)p(x) \s .
\eeq
To this end, we will use the fact that at fixed $m$, the probability that the system has local
energy $E$ is given by
\beq
p(E)=Z_E/Z_m
\label{eq:pe}
\eeq
where
\beq
Z_m=\int_m d\Es \Omega_S(\Es)e^{\Sr(\Er)}
\eeq
i.e. $Z_m$ is the total number of states at fixed total energy $m$.
Clearly, the probability that the local energy is $E$ is just given by
the total number of states which have local energy $E$ divided by the
total number of states $Z_m$.

We therefore find, that in the microcanonical ensemble, the probability
distribution of \S\ in contact with a large reservoir is given
by
\beq
p(\Es)d\Es=\sum_E \Omega_S(\Es)\Omega_R(E) e^{-\Es\beta_E}d\Es/Z_m
\label{eq:sume}
\eeq
where the sum is taken over all $E$ consistent with total energy $m$.
We see that one does not recover the usual thermal distribution.  Rather,
one has a probability distribution of thermal distributions.
There is also no decoupling of the probability distribution of the system
from the reservoir.  In other words, one does not obtain a simple
probability distribution like Eq. (\ref{eq:canonprop}) which does not
depend on the reservoir.

Although the microcanonical distribution is appropriate for an isolated
system, 
there may be situations where the microlocal distribution is also
appropriate. Such situations include cases where superselection rules
single out a particular $E$ (for example charge conservation or angular
momentum conservation may not allow transitions from one value of $E$ to
another).  One also may have cases where there is a large gap between
various values of $E$ so that once the system takes on a particular value
of $E$ it is unlikely to change, as this would require a large random
fluctuation.  In such cases, a smaller system in contact with a large one
would behave as if it were in a canonical ensemble.

Finally, we note that the local temperature, as we have defined it,
is a function of $E$.  It is for this reason that we have explicitly
put in this dependence by writing $\beta_E$.  There will be different ``temperatures'' dependent
on what value of $E$ the entire system is found in.

\subsection{Global temperature and conserved energy}
\label{ss:global}

As we saw in the previous subsection, a small system interacting with a
reservoir behaves as if it is at a number of different local temperatures
$\beta_E$.
Eq. (\ref{eq:sume}) gives the probability distribution 
in terms of this local temperature, and the local
energy $\Es$.  However, the local energy is not a conserved quantity, 
and it is not the energy that an observer
will ascribe to \S, since it does not contain the interacting term.  
We therefore define the conserved energy of a system interacting with
another system \R\ as
$\Eo\equiv m(\Es,\Er)-m(0,\Er)\simeq \Es\frac{\partial m}{\partial \Es}$.  
This is the change in the conserved
energy of \S, if one only makes changes to its energy levels.  
Clearly, $\Eo$ is also a function of $E$ but we will not write this
explicitly.
We will now show that $\Eo$ gives
the energy levels of \S\ in the presence of \R.
Using the definition above we can rewrite Equation (\ref{eq:sume}) as
\beq
p(\Es)d\Es=\sum_E \Omega_S(\Es)\Omega_R(E)e^{-\Eo\beta_o(m)}d\Es/Z_m \s.
\label{eq:canono}
\eeq
where the global temperature is defined as
\beqa
\beta_o&\equiv&\frac{\partial \Sr(m)}{\partial m} \nonumber\\
&=&\frac{\partial E}{\partial m}\beta_E
\label{eq:globalt}
\eeqa
Note that this global temperature does not depend on $E$.
We have the important relationship
\beq
\beta_E \Es=\beta_o\Eo \s .
\label{eq:importantrelation}
\eeq

$\Eo$ can be thought of as the effective energy.  
I.e. it is the energy of the system in the presence of the
reservoir (in general, it is the energy of the system in the 
presence of its interaction with another system).  $\beta_o$
can be thought of as the closest thing one has to a physical temperature.
If this is not yet clear from the above definitions, it should become clearer 
when we examine the example below.

\subsection{An example: physical significance of global quantities $\Eo$ and $\beta_o$}
\label{ss:physsigofglobal}
We do not have a single thermal distribution, but rather
a probability distribution of canonical ensembles, one for each accessible
$E$.  However, the quantity $\beta_o$ is the same for each ensemble.
In Subsection \ref{ss:unequal} we will see that $\beta_o$ does possess a 
crucial quality of temperature -- namely, two systems in equilibrium will
have the same $\beta_o$.  Here, we point out some other physical properties
that the global quantities $\beta_o$ and $\Eo$ have. 

Let us examine the physical significance of $\Eo$ when we look at the spin model
of equation (\ref{eq:twoissings}).  Let us imagine that
we have a homogeneous system, so that $h_1=h_2$, and $J_1=J_2=J_{12}$, 
and we can therefore drop the subscript.  Let us also work in the
microlocal ensemble with fixed $E$ for the time-being as we simply want to
understand the significance of $\Eo$ and $\beta_o$.
Let us consider System 1 to be a single spin (we will henceforth treat
System 1 as the small system \S\ and System 2 as the reservoir \R).  
This single spin acts like a probe, and can be thought of as a thermometer.
Then inserting Eq. (\ref{eq:spinentropy}) into Eq. (\ref{eq:canon}), one can verify that the probability
of the thermometer being spin up (and hence, having local energy $h$), is equal to
$(N+\e)/2N$ which is exactly as one expects, since this is just the
fraction of spins which are up in the entire system (here, $N$ is the total
number of lattice sites).  
However, the true energy levels of the spin are not $\pm h$
as they would be if the system was non-interacting.  
One must also add the field due to all the spins in System
2 (i.e. the rest of the system, not including the probe spin). 
This means that the single spin actually feels a magnetic field of $h-Je$.  
This is exactly equal
to $E_o$.
Therefore, someone measuring the energy levels of the thermometer
would conclude that the thermometer had energy levels $\Eo$
and that based on its probability distribution
it is in a thermal distribution at temperature $\beta_o$.  
This is exactly what is given by Eq. (\ref{eq:canono}).
$\beta_o$ is therefore, the physically significant temperature from the point of view 
of this type of a  determination
of temperature.  Note however, that because of the self-interactions in the system, $m\neq N \Eo$.  
I.e. one cannot
find the total energy by adding up all the locally conserved energies of each spin.
This is just another manifestation of the non-extensivity of the system.

\subsection{Physical significance of the local quantities $E$ and $\beta_E$}
\label{ss:physsigoflocal}

The temperature of the probe in the above example was $\beta_o$, but we
will now show how to use a 
probe to measure the local temperature $\beta$.
From the point of view of the formalism, the local temperature is a useful
quantity. It is clearly an intensive quantity, as can be seen from Eq.
(\ref{eq:additive}).  One the other hand, the global temperature is not
an intensive quantity.  Doubling the size of the system along with the total
energy, will result in a change in the temperature, as can be seen simply
from its definition (we will see encounter this more explicitly in \ref{ss:unequal}). 
We will also find that the local temperature has a very
strong physical significance in general relativity, where it is the
physical temperature measurable by free falling observers.  One would
therefore like to know how to measure it in other theories. 
It can be measured using the following method.  We again
use the example from the previous subsection  of a single spin probe 
interacting with a larger system and identical to it. 

We use the probe to measure the temperature by first slowly and 
adiabatically drawing the probe away 
from the rest of the spins
until it no longer interacts with the system (i.e. $J_{12}=0$).  Its energy levels
then become $\pm h$.  From Eq. (\ref{eq:importantrelation}), one sees that this 
spin then acts like a thermometer 
measuring the
local temperature $\beta_E$.  I.e. in an adiabatic change the state of the 
system remains constant and its average spin is still
$e/N$ .  However, the energy levels we would ascribe to the spin are now
different.  We would no longer say that the energy levels are given by
$\Eo$ since the magnetic field is now zero and the $eJ$ term no longer
contributes.  As a result, the spin acts as if it is in a thermal bath of
temperature $\beta_E$ and has energy levels $\pm h$.

One can use an almost identical method which reminds one of the 
measurement made by a freely falling observer in general relativity. 
In general relativity, the local temperature is in fact the physical temperature as measured
by an individual who is freely falling.  The global temperature $\beta_o$
is the temperature measured by an individual at infinity.  With a spin
system, there is no distinction between different observers.  However,
an analogy with general relativity motivates the following method.
One slowly (adiabatically) applies a local magnetic field $B=Je$ to the probe spin.  This is exactly
the magnetic field which cancels the magnetic field of the surrounding
spins which are acting on the probe.  The probe spin then has energy levels
of $\pm h$ and since its state has not changed, it is as if it is in a heat
bath of temperature $\beta_E$.  One can think of the applied field as being
analogous to the gravitational ``force'' which gets canceled when
one goes into free fall.  

It is worth noting that there is a significant different between
the method for measuring the local temperature in the case we have
described, and the case where the interactions are short-range (in the Ising model
say).  For example, when we remove a single spin from this system, we are
moving it in a known potential, since the interaction is mostly due to the entire system
as a whole.  On the other hand, in the Ising model,
where the potential is due to nearest neighbor interactions, one does not
know the local potential, since it is random.  Therefore, in the long-range
case, one can liberate all the long-range interaction energy by removing a spin.
In the Ising model, much of the interaction energy is contained in 
thermal fluctuations, and cannot all be liberated in such a manner.

\subsection{Unequal local temperatures at equilibrium, equal global
  temperature}
\label{ss:unequal}

It is a standard result that for two non-interacting systems,
the temperatures will be equal when they are brought into
equilibrium.  
We now extend these results by showing that 
the local temperature of two systems in 
thermal contact, need not be
equal, while each system's global temperature will be equal.
This in many respects justifies calling $\beta_o$ a temperature,
even though it is not the temperature
in the usual sense, since the microcanonical ensemble does not
lead to a canonical ensemble.

We allow the two subsystems to exchange energy but keep the total
energy $m$ fixed, while the local energy $E$ need not be fixed.
At equilibrium, the system will be found in the most probable configuration.
I.e. the entropy will be an extremum so that the system is in the macroscopic state with
the most number of microstates.
We can then find the extremum by varying $E_1$ and $E_2$ at fixed $m$. 
The entropy of two systems is given by Eq. (\ref{eq:additive}), and we  now
find the
extremum to give the most probable configuration.

\beqa
dS & = & \left(\frac{\partial S_1}{\partial E_1}\right)_m dE_1 +   
\left(\frac{\partial S_2}{\partial E_2}\right)_m dE_2 \nonumber\\
&=& \left[  \frac{\partial S_1}{\partial E_1} \frac{\partial E_1}{\partial E_2}
  +   \frac{\partial S_2}{\partial E_2}\right]_m dE_2 \nonumber\\
&=& 0
\eeqa
using the definition of local temperature of Eq. (\ref{eq:localt}),
we find
\beqa
\beta_2&=&- \beta_1   \left(\frac{\partial E_2}{\partial E_1}\right)_m \nonumber\\
&=& 
\beta_1
\frac{1 +   \frac{\partial G(E_1,E_2)}{\partial
      E_2}}
{1 +   \frac{\partial G(E_1,E_2)}{\partial
      E_1} }
\label{eq:gentempdif} \label{eq:tempint}
\eeqa
where $\beta_i$ is the local temperature of each system, and the dependence
on $E$ is implied.  One can also easily verify that the global temperatures 
of each system are equal, just by using Eq. (\ref{eq:tempint}) and the
definition of $\beta_o$  This to an extent justifies the term
{\it temperature}. 
Note
however, that if we move two systems together which are both initially at
the same global temperature $\beta_o$, then their new equilibrium temperature
can be at a different global temperature $\beta_o'$.  This is due to the
addition of new coupling terms in the total energy.  The global temperature
is therefore not an intensive quantity.  We will see this in the example
below.

Using the two coupled Ising models of Eq. (\ref{eq:twoissings}) 
we would get a temperature difference
of
\beq
\beta_2
= \beta_1 \frac{1-J_{12}e_1/h_2-J_2e_2/h_2}{1-J_{12}e_2/h_1-J_1e_1/h_1} \s .
\label{eq:isingtempdif}
\eeq
One can solve for $\e_1$ and $\e_2$ as a function of the total spin excess $e$ 
by explicitly calculating the temperature using 
Eq. (\ref{eq:spinentropy})
and equating it with the equation above.  
%
Solving these two equations for $e_1(e_2)$, will then allow us
to give an expression for the temperature difference.
\beq
\frac{\partial S_1}{\partial E_1}/\frac{\partial S_2}{\partial E_2}
= \frac{1-J_{12}e_1/h_2-J_2e_2/h_2}{1-J_{12}e_2/h_1-J_1e_1/h_1} \s .
\label{eq:isingtempdifwithe}
\eeq

Using Equation  (\ref{eq:spinentropy}) for the entropy
of the i'th system gives
\beq
\frac{\partial S_i}{\partial E_i}
=
\frac{1}{2h_i}\log{\frac{N_i-e_i}{N_i+e_i}}
\eeq

This gives the following equation
\beq
\left(\frac{N_1-e_1}{N_1+e_1}\right)^{h_2^2(h_1-J_{12}e_2-J_1e_1)}
=
\left(\frac{N_2-e_2}{N_2+e_2}\right)^{h_1^2(h_2-J_{12}e_1-J_2e_2)}
\eeq
which can be solved graphically for $e_1(e_2)$.  This can then
be substituted back into Equation (\ref{eq:isingtempdif}).  We 
shall not do so here.

Now if initially these two systems (or clusters),
are far apart, and at
equal global temperature, then when pushing them together one
cannot do so both adiabatically and isothermally (constant global
temperature) as one can do in the non-interacting case. This can
be seen from Eq. (\ref{eq:importantrelation}).  Moving the
systems together adiabatically requires keeping $\Eo\beta_o$
fixed.  But since $\Eo$ changes when $J_{12}$ becomes
significant, one cannot keep $\beta_o$ constant.  By recalculating $\Eo$
one can therefore calculate the new global temperature. We see therefore
that the global
temperature is not an intensive quantity.

Finally, one can consider what happens when one is in the grand-canonical
ensemble.  I.e. we allow the number of particles $N_1$ and $N_2$ to
change, while keeping the total number of particles $N$ and volume $V$ fixed.
In this case, one can define the local chemical potential in the
same way as we defined the local temperature
\beq
\mu=-T\left(\frac{\partial S}{\partial N}\right)_{E,V}
\eeq
and one finds that
\beq
\mu_1\beta_1=\mu_2\beta_2 \s .
\label{eq:globalchempot}
\eeq

This leads one to see that the local chemical potentials of two systems
will also not be equal, and that the ratio between the two chemical
potentials is the inverse of Eq. (\ref{eq:gentempdif}).  One can
likewise define a global chemical potential 
\beq
\mu_o=\frac{\partial m}{\partial E}\mu_E \s .
\eeq

Finally,
the preceding discussion allows us to write the average entropy of 
two systems.
Previously, we wrote the entropy of two systems as $S(E_1,E_2)$, as in
Equation (\ref{eq:additive}).  I.e. we
gave the entropy of the two systems when one had energy $E_1$ and the other
$E_2$.  However, if the two systems are in contact, the energies will
fluctuate until the system is in its most probable configuration.  We can
therefore write the average entropy of the two systems as
\beq
S(E)=\bar{S}(E_1,E_2)
\eeq
where it is understood that this is the average entropy of the combined
system.

\section{An example: the long range lattice model}
\label{sec:exising}

In order to better understand the points raised in the preceding sections,
it will be useful to work out a very simple example.  We will consider 
a single system of spins interacting view a long range potential which is constant.
This model is simpler than the one of Eq. (\ref{eq:twoissings}), and 
has total energy
\beq
m=h e -Je^2/2 
\label{eq:oneising}
\eeq
where we have simply dropped the subscripts from Eq. (\ref{eq:twoissings})
and gotten rid of the second cluster.

We essentially work in the microcanonical ensemble, and use the formalism we
have introduced.  There are other long-range lattice models which attempt
to solve similar interactions such as the 
Curie-Weiss model
where the interaction is made to scale
inversely to the number of lattice sites.  The dependence of the
interaction on the size of the system is problematic, but it ensures that the
thermodynamic limit exists.  A generalization of this is the Kac model\cite{kac}
which has an interaction with an exponential cut-off.
Here, because of our formalism, there is no need to introduce such a cut-off.
The model we explore is likewise related to the mean field
approximation of the Ising model, although in the mean field approximation,
it is as if one is
working in the microlocal ensemble, rather than the microcanonical ensemble
as we do here. In other words, in the mean field approximation, one will
not see the effects of having multiple values of $E$.
Another model which has been extensively studied are those with
$1/r$ potentials (see \cite{luijten-blote} for a recent review).  

In Equation (\ref{eq:oneising}) the local energy is $E=he$,
and for fixed $m$ there are two possible values of the local energy which
can be obtained by solving Eq. (\ref{eq:oneising}) for $e$ 
\beq
e_\pm(m)=\frac{h}{J}(1\pm k(m))
\eeq
where $k$ is 
\beq
k=\sqrt{1-\frac{2Jm}{h^2}} \s .
\label{eq:k}
\eeq
It is these two values of $e$ which will give us the two different local
temperatures.
From Eq. (\ref{eq:globalt}) and (\ref{eq:oneising}) we get the 
relationship between the global and local temperatures
\beqa
\beta(e_\pm)&=&\beta_o(1-\frac{Je_\pm}{h})\nonumber\\
&=&\mp\beta_o\sqrt{1-\frac{2mJ}{h^2}} \s
\label{eq:schwar-esque}
\eeqa
and likewise, using $E_o=E\partial m/\partial E$
we get
\beqa
E_o&=&(1-Je_\pm/h)he_\pm \nonumber\\
&=&\mp k(m) E
\label{eq:globale}
\eeqa
We will henceforth use $\beta_\pm$ to represent the two local temperatures.
It is also worth noting that Equation (\ref{eq:globale}) for the
conserved energy $E_o$, gives the energy levels 
of a spin in the presence of the effective magnetic field due to
the applied field $h$ and all the other spins of the system.
In other words, $E$ is the raw energy levels $\pm h$ in the absence
of the long range interactions, while $\Eo$ gives the energy levels
in the presence of interactions.

One can also arrive at Equation (\ref{eq:schwar-esque}) using the following
method which is highly illustrative.  Consider the example in the previous
section of two spin systems with a local temperature difference
given by Eq. (\ref{eq:isingtempdif}).  Now imagine that System 2 is being used as a 
thermometer to measure the temperature of System 1.  
I.e. System 2 has no long range interactions and 
minimal energy $J_2=J_{12}=0$. In this case, Eq. (\ref{eq:isingtempdif})
gives
\beq
\beta_2
= \frac{\beta_1}{1-J_1e_1/h_1} \s .
\label{eq:probenoint}
\eeq
This is precisely the same relation as Eq. (\ref{eq:schwar-esque}) giving
the relationship between the local and global temperatures.  We can conclude
from this that a spin which does not have long range interactions with the 
rest of the system will ``feel'' the global temperature -- its local
temperature will be the system's global temperature. On the other hand,
a system which is identical to the rest of the system will obviously have the
same local temperature as the rest of the system. 

However, as we know, this temperature depends on which branch of local
energy the system is in i.e. whether it is in the state $e_+$
or $e_-$.  If one were to look at a single spin in 
order to determine
the temperature, one would not find the spin in a thermal state.  Rather,
the spin would be in a distribution given by Eq. (\ref{eq:canono}).  I.e.
it is in a distribution of two possible canonical ensembles, with two
local temperatures $\beta_\pm$ corresponding to spin excesses of $e_\pm$. 
The conditional probability of a single spin being up, given a spin excess of
either $e_\pm$ is given by Eq. (\ref{eq:canon}).  I.e. the
probability of it being spin up having local energy $h$ is
\beq
p(h|e_\pm)=\frac{e^{\beta_\pm h}}{e^{\beta_\pm h}+e^{-\beta_\pm h}}
\eeq
The probability of the system being in the state $e_+$ or $e_-$ is given by
Eq. (\ref{eq:pe})
\beq
p(e_\pm)=\frac{\Omega_\pm\cosh(\beta_\pm h)}{\Omega_-\cosh(\beta_-h)
  +\Omega_+\cosh(\beta_+h)}
\eeq
where $\Omega_\pm$ is the number of states of the total system with spin
excess $e_\pm$.  It is given by 
\beq
\Omega_\pm=\frac{N!}{(\frac{N+e_\pm}{2})!(\frac{N-e_\pm}{2})!}
\eeq
and can be approximated by Stirling's equation.
The total probability of a spin being up or down is then given by Eq. (\ref{eq:canono}).
\beq
p(\pm h)=\frac{\Omega_+e^{\pm\beta_+h}+\Omega_-e^{\pm\beta_-h}}{2\Omega_-\cosh(\beta_-h)
  +2\Omega_+\cosh(\beta_+h)}
\eeq 
There is nothing special about the particular  spin we are using as a probe, 
and so the average orientation $p(h)-p(-h)$ should be equal to the
average magnetization of the entire system $\e/N$.
We therefore have
\beq
e(\beta_o)=N\frac{\Omega_-\sinh(\beta_-h)+\Omega_+\sinh(\beta_+h)}{\Omega_-\cosh(\beta_-h)
  +\Omega_+\cosh(\beta_+h)}
\eeq
while in the microlocal ensemble (fixed $e_\pm$), we have
\beq
e_\pm(\beta_o)=N\tanh(\beta_\pm h)\s.
\eeq

Finally, we note that one finds the usual phase transitions at $\beta_o=J/2$.
Here however, the phase-transition is real, not like the false phase
transition that can occur in the mean field approximation.  It therefore
exists in 1 dimension.  It will also have the property that there will
be different average $e$, depending on which microlocal ensemble the system
is in.

%

\section{A black-hole analog}
\label{sec:bhanalog}
\subsection{Lattice model as a black-hole analog}

In general relativity, one encounters a number of interesting thermodynamical
effects.  Perhaps the most well known is the Tolman relation\cite{tolman}.
One finds, that in curved space, the temperature, as measured by local
free falling observers varies from point to point.   This is usually
interpreted as being due to the red-shifting of frequencies due to the 
curvature of space-time.  We have already seen that for other systems
with long-range interactions, one has a variation of the local temperature.
The results are closely related, and here we will see that they can have
exactly the same form.  This gives a new interpretation to the Tolman
relation;  it is not the sole domain of general relativity as is usually
believed, but instead arises
in other theories with long-range interactions.  However, what makes
the variation of local temperature special in general relativity, is
the physical meaning it has.  We saw that
in the lattice
model, the local temperature was not a constant throughout the system
(just as in general relativity) -- however, the local temperature of the
lattice did not have the same physical interpretation as it does in general
relativity (where it is the actual temperature as far as free-falling
observers are concerned).  In the lattice model, the local temperature
was only the physical temperature if we applied a local magnetic field
to cancel the effective magnetic field due to the rest of the system.
This is closely related however, since the applied magnetic field is
very similar to the effect of free-fall in general relativity.

Another important thermodynamical phenomenon in general relativity is
the black-hole entropy.  This is often viewed as the key to understanding
quantum gravity, since any theory of quantum gravity should presumably predict the
correct value of the black-hole entropy.  It is therefor important
to understand what aspects of the black-hole entropy are specifically
related  to gravity, and what aspects arise in other theories.  We are
therefore motivated to explore the similarities of our lattice model,
with the types of effects one finds in general relativity.

Inspecting Equation (\ref{eq:schwar-esque}) for the lattice model, 
one can't help but be struck
by its similarity with the Tolman relation from general relativity.  
Indeed, defining $r=h^2$ and
putting the coupling $J=G$ where $G$ is the gravitational coupling constant,
we see that Equation (\ref{eq:schwar-esque}) becomes
\beq
\beta_\pm
=\mp\beta_o\sqrt{1-\frac{2Gm}{r}} \s.
\label{eq:schwar}
\eeq
The positive solution is exactly the Tolman relation
for the red-shifting of temperature in the Schwarzschild geometry  
-- the Schwarzschild geometry being the space-time of an uncharged
non-rotating black-hole or 
spherically symmetric star.  In Appendix \ref{ap:newtonian}
we will see that one obtains similar red-shifting effects
in a theory with a Newtonian potential.

Let us now show that our Ising model can be thought of as
a black-hole analog.
For $r>2Gm$ Eq. (\ref{eq:schwar}) behaves like the exterior Schwarzschild solution
(the solution outside of the black-hole).
As we decrease $r$, the local temperature becomes hotter and hotter 
(for fixed global temperature).

Setting $r= 2Gm$ gives the black-hole analog (or perhaps more appropriately, the point
in space where an observer would be on the horizon).  
In this case, there is an infinite "red-shifting" between
the global temperature and the local temperature as can be seen from
Equation (\ref{eq:schwar}).  This ``black-hole'' solution is not only special
because of the divergence in the red-shift -- it is also the point where the
two solutions $e_+$ and $e_-$ coincide.  There is therefore only one local
temperature.  One can therefore see that it is only the black-hole analogue solution
which is thermal.  All other solutions do not give a thermal distribution.
The black-hole solution is also special in that there is a degeneracy in
the energy levels--
a point which we will become important when we discuss the system's entropy.

There are two interesting cases to consider:
(1) the case where the global temperature $\beta_o$ is finite, and (2) where 
the global temperature is 0.

In case (1), if the global temperature is finite, then at the point that
$r=2Gm$
the local temperature diverges.
This is explained in the spin model by inspecting Eq. (\ref{eq:globale}) 
and noting that the point $h^2= 2Jm$ (the analog of the black-hole horizon) 
corresponds to $h=J e$.  This is exactly the point $E_o=0$ and 
therefore an individual 
spin sees no effective magnetic field.  In
other words, both the energy levels of a single spin are zero.
Therefore, since $\beta E=\beta_oE_o$ is finite
for finite $\beta_o$ we must have an infinite local temperature.  Also,
since individual spins see no net magnetic field, there is no preferred
spin direction, and one finds $e=0$.  This solution is therefore the solution
with maximal entropy.

The second case, with zero global temperature, can be thought of as being analogous to the
extremal black-hole (i.e. the charged black-hole solution with zero temperature).  
In this case, we can have a finite local temperature.  This again comes from
the relation  $\beta E=\beta_oE_o$ and the fact that $\Eo=0$.  In this
case, $e$ is more or less arbitrary.

The gravitational analog has another interesting property.  In gravity, one
requires that the radius of the black-hole is at $R=2Gm$ which is the
so-called Schwarzschild radius, one cannot have $R<2Gm$.  The same is true
here.  There is no value of $e$ which would allow $h^2<2Jm$.  The black-hole
analog solution occurs at the maximum of energy $m=h^2/2J$ which is exactly
the Schwarzschild radius.  The limit $h^2=2Jm$ therefore corresponds to a
horizon in a very real sense.  One can of course have negative $J$, in contrast
to physical black hole solutions.  In the case of lattice spins, the solution in
this case is the stable one.


The case of $e_-$ reminds one of the situation interior to a black hole,
since in this case, the local temperature is negative as can be seen from
Eq. (\ref{eq:schwar}).  Likewise, inside the black-hole, the light-cone
is tilted over, in such a way that energies which are positive outside
the hole, are negative inside.
In the black-hole analog case, the negative local temperature
can be understood from Eq. (\ref{eq:globale})
by noting that positive conserved energies $E_o$
correspond to negative local energies.
In other words, a spin is more likely to be pointing up with energy $h$,
even though this corresponds to a greater local energy (it instead
corresponds to a smaller conserved energy).

We now turn our attention to the entropy of these solutions.
This is of interest because the black-hole entropy is usually 
considered to have some unique properties:  (1) the black-hole's
entropy is proportional to its area and not its volume.  In the
case of three spatial dimensions, this means that the black-hole
entropy is proportional to the square of its total energy (since
the black-hole radius is at $2m$).  (2) the black-hole entropy is 
universal - the constant
of proportionality between the entropy and the area is the same
regardless of the past history of the black hole.  This means
that no matter what type of initial matter formed the black hole,
its final entropy will only depend on the total energy of the black-hole
(or other conserved quantities in the case of charged and rotating
black-holes).
(3) before a system forms a black-hole, its gravitational entropy
is zero, while the black-hole entropy (which is enormous) appears
suddenly, when the system forms a black-hole.  The system may
have some material entropy before it forms the black-hole, but
this is negligible compared to the sudden increase in entropy
it gets when it forms a black hole.
The particular spin model under discussion does not possess 
properties which are identical to the black-hole.  However,
it does possess similar characteristics to the three mentioned
above.

To aid in this discussion, it is worthwhile to add the constant terms
back into our expression for the total energy $m$ of the spin model.  
From Eq. (\ref{eq:ising}) one can
put back the constants, so that instead of Eq. (\ref{eq:oneising}) one
gets
\beq
m=he^2-Je^2/2 -JN/2
\label{eq:oneisingwithn}
\eeq
where $N$ is the total number of spins.

Now the entropy one finds will depend on what one specifies
about the system.  Consider the case when one knows not only
$m$, but also the energy $E_o$ of every spin.  In this case,
the entropy of the system is zero, since knowing the energy
$E_o$ of each spin is the same as knowing the spin itself,
so one has complete knowledge of the system.  However,
when the system becomes a black-hole analog at $h^2=2Jm$
the energy levels $E_o$ of each spin
become doubly degenerate and the system suddenly acquires
an entropy of $\log 2$ per spin.  This then is similar
to property (3) of a black hole.  Such a property also exists
if one doesn't know the energy of each spin, but instead knows
the total local energy of the system.  I.e. one knows whether
the system is in the $e_+$ state, or the $e_-$ state.  When
the system forms the black-hole analog, these two different
states merge, and one acquires an additional (although
negligible) entropy of $\log  2$.  However, one interesting
property of this entropy $\log 2$, is that it is only a function
of the form of the interaction, and not of the particular system.
In this system, we have a factor of $\log 2$ because the potential
is quadratic, and there are two possible local energies $he_\pm$
for fixed total energy $m$.
It is tempting to regard this as a type of
universality, similar to property (2) of the black-hole.  The factor
of $\log 2$ comes because of the form of the interaction, and has
nothing to do with the particular system, just as the black-hole
entropy comes from the gravitational interaction and has nothing
to do with the particular system.
For a general potential, there will be $n$ possible local energies
at fixed energy and one might
regard $\log n$ as being the entropy associated with the interaction.
One can also make the degeneracy arbitrarily large, by considering
higher level spins, rather than just the two level systems we have
been considering here.

Note that the degeneracy in $E_o$ which occurs in the spin model
also has a counterpart in the black-hole.  There, one also finds
that the conserved energy is zero on the horizon.  In some sense,
this is what enables one to ``pack'' a large amount of entropy at
no energy cost, close to the horizon.

Finally, one can ask about the non-extensive properties of the
entropy.  The entropy is non-extensive
in the sense that it doesn't scale proportionally
with the total energy $m$.  In other words, the entropy
is proportional to the local energy $E$ (and $N$), but because 
the total energy $m$ does not scale linearly with $E$ and $N$,
the entropy will not be a linear function of the total energy.

If one uses the entropy as given by Eq. (\ref{eq:spinentropy})
then the entropy will scale as both $N$ and $e$.  
If one uses the definition of entropy
discussed above, then the entropy will scale with $N$.
On the other hand, the scaling of the total energy is given by 
Eq. (\ref{eq:oneisingwithn}) so a general system will not
have an entropy which scales like $m$.  For the case of the black-hole
solution i.e.  $h^2=2Jm$, the total energy is
\beq
m=\frac{J(e^2-N)}{2}
\eeq
which has a completely non-extensive part (scaling-like $e^2$)
as well as the extensive part (scaling like $N$).  

Here, the entropy, while exhibiting non-extensivity as a function
of total energy, does not scale the same way as a black-hole
(i.e. $S\propto m^2$). 
However, one can imagine easily constructing an interaction which has an
entropy which has the same dependence on $m$ as the black hole.
For example, by having an interaction
of the form $m\propto \sqrt{E}$.  Then, for a locally extensive
system (i.e. $S\propto E$) such as the spin models we have been
considering, one will find the same entropy scaling behavior
as a black-hole.


We will discuss in more detail in Section \ref{sec:scaling} how
one can derive the scaling relations for the entropy based on the
considerations introduced here.

Finally, we note, that one can arrange the phase transition so that it puts
the system at $e=0$ and one finds a second order phase transition into the black-hole
analog solution.  This is in contrast to the black-hole case, where the jump
in entropy suggests a first order phase transition.  However, if one looks
at the entropy given that one knows the value of $E_o$ for each spin, then
the analog does indeed have a discontinuity in entropy.

\section{Non-extensive scaling laws}
\label{sec:scaling}

As discussed in Section \ref{sec:bhanalog}, the scaling of the entropy will
no longer be purely extensive.  Here, we will show how to quantify the
degree of non-extensivity for particular systems.  The main idea is to
use the principle of local extensivity which gave as Eq. (\ref{eq:additive}).
In other words, in terms of the local energy $E$, the entropy is an extensive
quantity.  This can be written as
\beq
S(\lambda E)=\lambda S(E)
\label{eq:entropyextensive}
\eeq
in terms of the total energy $m$, the system will not be extensive.
Here, we will work in the microlocal ensemble - in the end, one
must sum over all $E$ consistent with $m$.  We will therefore in this section
write $S$ as a function of $E$ to remind ourselves of this.  In the
case of densities, it is understood that $s$ is a function of  $\rho$.

We now use a second ingredient, which is that we use the 
the Gibbs-Duhem relation \cite{gd}
%
%
\beq
s=\beta(\rho+p)-\mu n \label{eq:gd-density}
\eeq
to relate the various thermodynamical quantities to each other.
Here, $s$,$\rho$,$n$ are the entropy density, energy density and particle number density
and $T$, $\mu$ and $p$ are the local temperature, chemical potential and
pressure.

It will prove easier (although not necessary), to multiply Equation
(\ref{eq:gd-density}) by a tiny volume element $V$ to get
\beq
E=T S(E)- pV +\mu N \label{eq:gd}\s .
\eeq
The standard derivation of the Gibbs-Duhem relation follows from the first law, and the
principle of local extensivity.  One has
\beq
dE=TdS-pdV - \mu dN \s .
\label{eq:firstlaw}
\eeq
Then, analogously to Equation (\ref{eq:entropyextensive}) one has
\beq
N(\lambda E)=\lambda N(E)
\label{eq:numberextensive}
\eeq
and also that in a small volume, the quantities $T$, $p$ and $\mu$
are intensive i.e. they do not change with $\lambda$.
One can then integrate the first law to obtain the Gibbs-Duhem relation.
This relation is known as an Euler relation
of homogeneity one.  Quantities which scale like $\lambda^a$
are Euler relations of homogeneity $a$.

We can now express the local energy $E$ as a function of 
total energy $m$, and also use the expressions Eq. (\ref{eq:globalt})
and Eq. (\ref{eq:globalchempot}) to express the local temperature
and chemical potential in terms of their global quantities $\beta_o$
and $\mu_o$.  Or, in the case of a continuum system, one can use
Eq. (\ref{eq:tvar}), Eq. (\ref{eq:muvar}) and Eq. (\ref{eq:pvar})
We would thus have all local quantities expressed in terms of global ones.

In the continuum case, one then gets for the entropy density
\beq
s(\rho)=\beta_o\frac{\partial \rho}{\partial \dot{m}}(\rho-p)-\mu_o\beta_o n
\eeq
which can then be integrated to give the total entropy in terms of global
quantities.  Just as we used $\Eo$, we here use the
conserved energy density $\rho_o$.  The equation
for $p(x)$ will depend on the the potential.  In the next section, we
will calculate this quantity for a gravitating perfect fluid, and
we will see that the entropy will not scale like the volume of the system,
but rather, approaches area scaling behavior as the system becomes more
strongly interacting.

As a general rule, one will obtain $S(m,\beta_o)$.  From this,
one can then calculate $S(\lambda m,\beta_o)$ in order to 
determine the scaling behavior of the entropy.  For
general interactions
one find that $S(\lambda m ,\beta_o)\neq\lambda S(m,\beta_o)$.
Instead, for homogeneous potentials, one finds  
\beq
S(\lambda m, \beta_o)=\lambda^a S(m,\beta_o)
\eeq
and the exponent $a$ then quantifies the degree of non-extensivity
of the system.

Perhaps the most famous example of this, is the case of a black-hole in three
spatial dimensions,
where, one finds the so-called Smarr
relation\cite{smarr}
\beq
S=\beta_o m /2 
\label{eq:smarr}
\eeq
which is an Euler relation of homogeneity $2$
in contrast to the non-interacting case of
$S=\beta E$ which is an Euler relation of homogeneity $1$.

One can show that the Smarr relation quantifies the
non-extensive nature of the interaction. 
Differentiating the Smarr relation Eq. (\ref{eq:smarr})
and applying the first law
$dm=T_o dS$  
one obtains
(in terms of a constant $\gamma$)
\beq
S =\gamma m^2 
\eeq
so that the entropy scales not as $m$ as it would in
the non-interacting case, but as $m^2$.  We had such a term
in our long range lattice model.
The black-hole's radius is at $R=2m$ and so in terms
of the black-hole area $A$ one obtains
\beq
S=\frac{\gamma}{16 \pi} A \label{eq:SofA}
\eeq
One likewise gets
\beq
T_o = (2\gamma m)^{-1} \label{eq:tofm}
\eeq
which is the correct expression (up to a constant of proportionality) 
for the Bekenstein-Hawking temperature.

Finally, it is worthwhile to explore some additional relationships one gets
for extensive systems.  Taking the derivative of the Gibbs-Duhem relation
Eq. (\ref{eq:gd}), and applying the first law Eq. (\ref{eq:firstlaw}),
gives
\beq
dTS=Vdp-Nd\mu
\eeq
which yields the following two relationships
\beq
\left(\frac{\partial T}{\partial p}\right)_\mu=\frac{V}{S}\s,\s\s\s 
\left(\frac{\partial T}{\partial \mu}\right)_\mu=\frac{N}{S} \s.
\eeq
These relationships between intensive and extensive variables
only hold for extensive systems, although they hold
locally for non-extensive systems.  


\section{Entropy scaling behavior in general relativity}
\label{sec:entropyfluid}

General relativity is another theory in which our assumptions of locality
and local-extensivity hold.  What's more, quantities like the local
temperature have a very real physical meaning -- the local temperature
is the physical temperature measured by an observer in free fall.
We will first discuss how our two assumptions hold in general relativity.
Then we will discuss the entropic scaling relations for the perfect 
fluid.  The principle motivation for the latter study comes from \cite{area},
where I showed that the entropy of a spherically symmetric material
(approximated as a densely packed set of shells) has an entropy which
is area scaling at the point before it forms a black-hole.
It is therefore seen that the area-scaling property of entropy
is not unique to the black-hole.  This suggests that this property
arises from the long range interactions of gravity, and is not
solely due to the horizon.  Here, we will see similar behavior,
however, because we can solve the equations exactly, we can
trace the entropy scaling behavior at all values of the gravitational
coupling constant.

Let us first see how our two assumptions hold in general relativity.
A review of thermodynamics in curved space can be found in \cite{mtw}.
Let us first consider the case where there is no gravitational
interactions.  The thermodynamical quantities, $\rho$, $n$, $T$, $\mu$,
$p$, and $s$ are taken to be the quantities measured in the rest frame
of the substance.  Let us now consider the case where we have
gravitational interactions.  In this case, we can go into the proper  
rest-frame of the material, and consider an observer
who is released into free-fall.  By the principle of equivalence,
this observer would measure the same quantities  $\rho$, $n$, $T$, $\mu$,
$p$, and $s$ (these are what we called the local variables).
An equation like the Gibbs-Duhem relation of Equation (\ref{eq:gd-density})
is a scalar equation. Since it holds in the non-gravitating case,
it also holds for the local free-falling observer.  Furthermore, since it is a
scalar equation, it holds for all observers.  

In general relativity, the local temperature 
$T=(\partial s/\partial\rho)^{-1}$ is a very real quantity,
as it is the temperature as measured by local free falling observers.
Likewise, the global temperature, $T_o$ is the temperature that would
be measured by an observer at infinity.  This corresponds almost exactly
to the case we were considering in the long-range lattice model.  There,
the global temperature could be measured
by isothermally taking a spin
and moving it away from the system so that it no longer felt the interaction,
and then measuring its temperature (this is like measuring the temperature
at infinity).  The local temperature could be measured
by canceling out the local magnetic field caused by the interaction, just
as going into free fall causes one to not feel the gravitational ``force''
(not including the tidal force).  Although in general relativity the local 
temperature is just as ``real'' as the global temperature, this cannot be used to create
a perpetual motion machine, because the energy one could extract by
moving from a hot local temperature to the cold temperature at
infinity is exactly canceled by the work needed to escape the
gravitational potential.  Gravity is universal i.e. all objects
feel it, so there is no heat engine that could be used to create a
perpetual motion.  In contrast, not all heat engines would feel the
spin-spin interaction which we introduced in the lattice model, however,
there, the local temperature did not have the same physical meaning
as it does in general relativity.  This is because the energy levels
of each spin are best described by the conserved energy $\Eo$ and not
the local energy $E$.  It is interesting that one
requires the equivalence principle in order for the
local temperature to be a real physical temperature.
On the other hand, if the local temperature is physical,
then one needs the universality of gravity 
in order to protect the second law of thermodynamics.  

We now turn to
the entropy scaling behaviour of the gravitating fluid.  We use the
Gibbs-Duhem relation to calculate the scaling behavior.
The actual  calculation, 
while instructive, is done in Appendix \ref{ap:scaling}.  We also describe
how to perform such calculations in greater generality 
in Appendix  \ref{ap:continuum}.

A related calculation to the one here is that of Zurek and Page\cite{zurekpage} who
have calculated the entropy (numerically) for
the case of a perfect fluid surrounding a black-hole, assuming a
specific equations of state.
 
For a spherically symmetric
fluid of constant density, one can calculate the entropy exactly,
and it is given by
\beq
S
= \frac{3k(R)R\beta_o}{4}
(1-\frac{N\mu{R}}{m})
\left[\sqrt{\frac{R}{2m}}\arcsin{\sqrt{\frac{2m}{R}}}-k(R)\right]
\eeq
where $R$ is the radius of the fluid, and 
$k$ is given by
\beq
k(R)=\sqrt{1-2m/R}
\label{eq:fluidentropy}
\eeq
and is virtually identical to Eq. (\ref{eq:k}).  $m$ is the total energy
of the material as measured at infinity (the ADM mass\cite{adm}), and
therefore we use the same symbol $m$ we have been using for the total energy.
Likewise $\beta_o$ is the temperature as measured at infinity, and is thus
the same quantity as we have been calling the global temperature.
Here, the gravitational constant $G$ has been set to $1$.

\begin{figure}
\centerline{
  \psfig{file=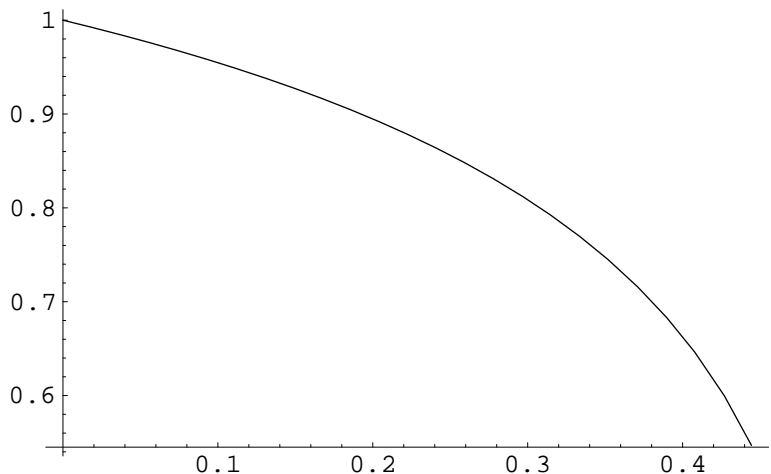}
  }
  \caption{$\gamma$ vs $m/R$ for a perfect fluid of constant density}
  \label{fig:fluidentropy}
\end{figure}

Earlier, we saw that for a black-hole, we had the Smarr relation $m=2T_oS$
while for ordinary matter, $m=TS$.  We showed  that the factor of 2 yielded
the area scaling property of the black-hole.  It is therefore interesting
to see how the entropy of the perfect fluid behaves.  Indeed, putting the
chemical potential to zero, we can calculate $\gamma\equiv ST_o/m$ as given by Equation 
(\ref{eq:fluidentropy}).  This is done in figure \ref{fig:fluidentropy}.
We essentially plot $\gamma$ versus the strength of the gravitational
interaction $m/R$.  We could have also put back the constant $G$ (in which
case one has $m/R\rightarrow Gm/R$) and plotted $\gamma$ vs $G$ holding
$m/R$ constant.

We find, that when the gravitational interaction is weak (i.e. $m/R$ is
small), the quantity $\gamma$ is $1$ just like in ordinary matter.  As we
increase the strength of the gravitational field, $\gamma$ gets smaller,
and approaches $1/2$ just as it would for a black-hole.  However, we cannot
plot $m/R$ greater than $4/9$, since at this point, the
central pressure diverges.  The strength of the interaction which
corresponds to a black-hole is $m/R=1/2$ (the Schwarzschild radius).
This can however be obtained if we have not only central pressure,
but also tangential pressure.  Indeed, we have done this for spherical
shells which have such tangential pressure, and seen that the matter
becomes area-scaling before a black-hole forms\cite{area}.  We
see therefore that while the system obeys the Gibbs-Duhem locally, it does not
obey it globally.
 This suggests that the fact that black-holes have an entropy proportional to their area
may be related to the long-range interactions of gravity rather than only
being a
special property of the horizon.  
Area scaling in gravitational systems exists even
though their is no black-hole horizon.

\begin{figure}
\centerline{
  \psfig{file=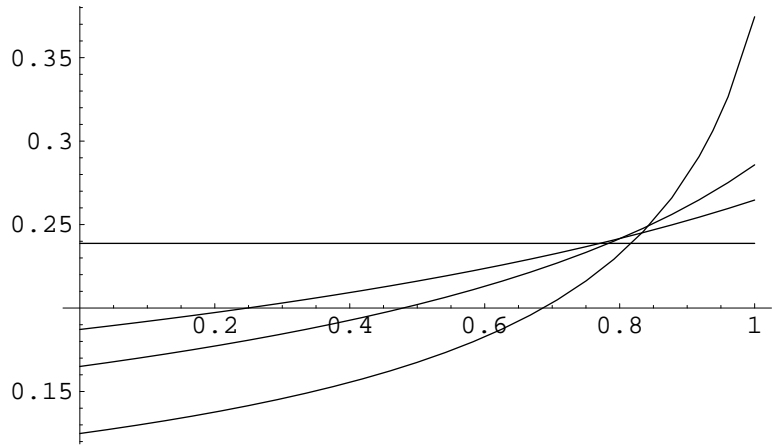}
  }
  \caption{Fraction of entropy density $s(r)/S$ vs. radius (normalized to $1$)
    for the perfect fluid.  The curves plotted are for a strength of
    gravitational interaction $m/R=4/9$,$1/3,$$1/4$, and $0$.  $m/R=0$
corresponds to the straight line, while $m/R=4/9$ corresponds to the
uppermost line}
  \label{fig:entropyvsr}
\end{figure}

The distribution of 
entropy is plotted in Figure \ref{fig:entropyvsr} for various strengths
of interaction.  When the gravitational interaction is negligible, the
entropy is constant throughout the sphere as one expects.  As the
strength of the gravitational coupling is increased, the entropy moves
to the surface of the sphere.  This intriguing effect helps explain why 
the entropy becomes more area-scaling.  In the case of tangential pressure, where one can
actually approach the black-hole radius, one finds that {\bf all} the entropy
lies at the surface of the material.

There is another remarkable property of the entropy of a perfect
fluid which is worth mentioning.  One should ask whether the
entropy as we have calculated, is an extreemum.  Indeed, it is,
however, it is only an extremuum if Einstein's equations are
satisfied\cite{joseph,thermgrav}.
It is remarkable, because there seems to be no apriori reason
why the entropy should only be an extremuum in curved space
for the particular space-time given by Einstein's equations.
This interesting connection is discussed in 
some detail in \cite{thermgrav}.

\section {conclusion}
\label{sec:conclusion}

We have introduced a formalism for studying the thermodynamics
of interacting systems.  The formalism is partly inspired from
our understanding of thermodynamics in general relativity.  This
allows us, not only to use general relativity to undertand
non-extensive thermodynamics, but also, one can learn more
about thermodynamics in curved-space by looking at 
thermodynamics in other interacting theories.

We have seen for example, that many of the properties
of black-hole entropy also exist in other systems.
Likewise, the red-shifting of temperatures has a place
in other theories of gravity which are not metric theories
and have a flat space-time.
One can therefore conclude that many of the effects
in general relativity have an analog in more classical
theories.  These results are helpful when attempting
to construct a quantum theory of gravity, because it
enables one to seperate the accidental aspects of black-hole
thermodynamics from the more fundamental ones.


We have also seen several new effects in non-extensive systems
which are worthy of more exploration.  We have found that
the local temperature can vary throughout a substance, and
also that an isolated system can appear to be in a distribution
of different temperatures i.e. canonical ensembles.
It would be interesting to apply this formalism to other theories.

Here, we have studied fairly simple systems, such as clusters of lattices
with different uniform long-range interactions.  We have generalized
the formalism for more complicated interactions, but it would be
useful to explore this further. In particular, one expects many
 related phenomena in other self-interacting
theories.  Non-abelian gauge theories such as $\phi^4$ theory may be
interesting arenas of study.  Numerical simulations might also be
particularly useful to study some of these effects in more complicated
systems.

It would also be useful to attempt to see these effects experimentally.
The case of two clusters of lattices might be realized by making the
clusters very small, so that the spacing between lattice sites is
much smaller than the range of the spin-spin coupling.

\appendix
\section{The definition of local temperature}
\label{ap:equalt}

In Section \ref{sec:longrange}\ref{ss:temp} we derived the distribution
of the microlocal ensemble by looking at a system \S\ and reservoir
\R\ in the
microcanonical ensemble.  
The local
temperature was defined as
\beq
\beta_E\equiv\frac{\partial \Sr(E)}{\partial E} \s .
\label{eq:localt3}
\eeq
i.e. it was defined in terms of the entropy of the reservoir.
In the non-interacting case, one tends to think of the
temperature as being 
\beq
\beta_E'\equiv\frac{\partial \Ss(\Es)}{\partial \Es} \s .
\label{eq:localt2}
\eeq
Our result of Section \ref{sec:longrange}\ref{ss:unequal}
show that $\beta_E\neq\beta_E'$.
The definitions are equivalent in the case where \S\ is just
a smaller part of a much larger system i.e. when we formally
divide a large system into \R\ and \S.
In this case, one can show
\beq
\beta_E = \beta_E' \s 
\label{eq:equivlocalt}
\eeq
because of symmetry.

To see this, we write the total energy of the total system as
\beq
m(E)=E+G(E)
\label{eq:selfint}
\eeq
where $E=\Er+\Es$ as before.  Then taking the partial derivative
of  Eq. (\ref{eq:selfint}) with respect to $\Es$ and holding $m$
fixed we find
\beqa
\left(\frac{\partial \Er}{\partial \Es}\right)_m
&=& -
\frac{1+ \frac{\partial G(E)}{\partial \Es}
}{1+ \frac{\partial G(E)}{\partial \Er} }
\nonumber\\
&=& -1
\eeqa
Combining this with Eq. (\ref{eq:tempint}) we obtain 
the desired result Eq. (\ref{eq:equivlocalt}).  

One important special case is when the reservoir has no long-range
interactions.
In this case, one has
\beq
\beta_E=
\beta_E'
\left[
1 
  +
    \frac{ \partial G(\Es)}
   {\partial \Es}
\right]\s .
\eeq
These results follow from Eq.
(\ref{eq:gentempdif}).  Note that in this case,
one also has that there is only one term in the sum in 
Equation (\ref{eq:canono}).  This is because in this case,
$\Er$ is uniquely determined from $\Er=m-\Es-G(\Es)$.
\section{A Tolman relation in Newtonian gravity}
\label{ap:newtonian}

We have seen in the case of the long range lattice model, that it has behavior reminiscent
of a Schwarzschild geometry.  This indicates, that many of the thermodynamic properties one associates
with general relativity, may be present in Newtonian gravity.  Indeed, we will now see that a Newtonian
type interaction does lead to the Tolman relation.  Here, we will see that it arises from the long range
interactions, and not necessarily from the curvature of space time.

We imagine that we have two gravitating systems with mass $M_1$ and $M_2$, 
and thermal energy $E_1$ and $E_2$ (which is the additional kinetic
energy present in the molecules of each system), and we imagine that their volumes are fixed.  
The systems are assumed to be a
distance $d$ apart but in thermal contact (one might imagine 
that there is a conducting wire connecting the two systems).  
We will consider the following Newtonian-type interaction
\begin{widetext}
  \beq
  m=M_1 + E_1 + M_2 + E_2
  -G_1(M_1 + E_1)^2
  -G_2(M_2 + E_2)^2
  - G_{12}(M_1 + E_1)(M_2 + E_2) \label{eq:interaction}
  \eeq
\end{widetext}
where the $G$ are coupling constants.  Note that the model uses the fact
that thermal energy also gravitates.
Using Equation (\ref{eq:gentempdif})
leads to a temperature ratio of
\beq
\frac{T_1}{T_2}
= \frac{1-G_{12}(M_2+E_2)-2G_1(M_1+E_1)}{1-G_{12}(M_1+E_1)-2G_2(M_2+E_2)} \label{eq:temps}
\eeq

Examples of systems which have such an interaction include two gravitating spheres separated by a
distance $d$, as well as two concentric spherical shells at radii $r_1$ and $r_2$.  In the latter
case, we have $G_i=G/2r_i$ and $G_{12}=G/r_2$ where $G$ is Newton's constant.  To first order
in the coupling constants, the temperature difference for two shells can then be written as
\beq
\left({\frac{T_1}{T_2}}\right)_{shells}
= 1-G(M_1+E_1)(\frac{1}{r_1}-\frac{1}{r_2})
\label{eq:tempsgr}
\eeq
One can use a completely independent method to calculate the temperature ratio in
full general relativity for this case, and one finds
that the results are identical for weakly interacting fields where general relativity and Newtonian
mechanics coincide.  
This, to a large extent, justifies the assumptions we made at the beginning of
this article.

Essentially, for gravity, the derived temperature difference coincides with what one expects from the
Tolman relation, except in this
case, there is a correction due to the fact that we are not considering a thermal system in a
fixed gravitational background, but rather, the thermal system is partly responsible for the gravitational
interaction.  For this reason, we see that the ratio does not only depend on the ratios of the redshifts
$1-G M_i/d$, but on $1-G(M_i+E_i)/d$.  I.e. the thermal energy $E_i$ also contributes to the redshift factor.

One can also argue that the local temperature difference is indeed real for
a freely falling observer, who essentially will be unaware of the
additional gravitational interaction.  Of course, this already invokes the
equivalence principle.
There are  
two other interesting points worth mentioning.  One is that in order to get
the temperature difference one needs to have a differences in charges (in
this case, a difference between $M_1$ and $M_2$).  It is this asymmetry
which is partly responsible for the temperature difference.  Additionally, 
one  needs self-interactions i.e. the thermal energy $E_i$ needs to also
gravitate.  Thus our model is not identical to Newtonian gravity, but
includes the fact that all energy gravitates.  This then seems to  be
the key ingredient which gives temperature differences.

\section{Continuum limit}
\label{ap:continuum}

In Section \ref{ss:unequal} we looked at the local temperature
and global temperature of two systems in equilibrium.  We then
examined a simple example of two clusters interacting via two different
uniform interactions.  It is worthwhile to generalize this.  For the
case of a small number of regions, one can use the methods introduced
earlier for just two regions.  However,  
one can imagine a more complicated interaction like one of Eq. (\ref{eq:ising})
where the interaction term is not a constant over any area, but instead changes from
site to site.  We can write instead $J_{ij}=J(x_{ij})$ and then write
all thermodynamics quantities as a function of the position $x_{ij}$.
In fact, it will prove  
simplest to go the case where we treat a system as a continuum -- it
is then easy to go back to the discreet case.

Let us now derive the relationships between the various local thermodynamical
quantities.  We essentially carry out a similar procedure as we did in
Section \ref{ss:unequal}.  In other words, we extremize the total entropy
at fixed total energy.  We consider the entropy in terms of a density
$s(\rho(x))$ where $\rho$ can be thought of being related to a spin
density.  One can think of $\rho$ as the local
energy density in analogy to the quantity $E$ i.e. it is the continuum
version of $E$.  
More explicitly, we can write $\rho(x)=h(x)\sigma(x)$
where $\sigma(x)$ is the spin at each site $x$, and $h(x)$ is the energy
gap at each site in the case where there is no interaction.
We will however, leave it general, and simply write $s(x)$ for simplicity
(with the understanding that $s$ is a function of the local energies).
I.e. 
\beq
S=\int s(x) dx \s.
\eeq

We then extremize this by taking the variation, and keeping $m$ fixed.
In order to do this, we append a constraint to the above expression,
so that we instead extremize
\beq
L=S+\lambda\left(m-\int \dot{m} dx\right) \s .
\eeq
Here, we are merely introducing the formalism.   Indeed $\dot{m}\equiv dm/dx$  may be a
complicated function, however, for the general lattice model of Eq.
(\ref{eq:ising}) $\dot{m}$ is a functional of the spins at each site.
It is in fact, also functional of $\rho$.
We can now vary $L$ with respect to 
$\delta\rho$. 
\beqa
\delta L &=& \int \left( \frac{\partial s}{\partial \rho}  - 
\lambda \frac{\partial \dot{m}}{\partial \rho} \right)\delta\rho\,\, dx
\nonumber\\
&=& \int\left( \beta(x)- \lambda \frac{\partial \dot{m}}{\partial
    \rho}\right)\delta \rho\,\, dx
\eeqa
Since this must vanish for all $\delta \rho$, we have that
\beq
\beta(x)= \frac{\partial \rho}{\partial\dot{m}}\beta_o
\label{eq:tvar}
\eeq
Here, we have set the constant $\lambda=\beta_o$.
One can see that in the case of no interactions, we have
\beq
m=\int \rho dx
\eeq
and therefore
\beq
\dot{m}=\rho
\eeq
so that we recover the standard result that the temperature is a constant.
One can likewise obtain
\beq
\mu(\rho)\frac{\partial \rho}{\partial\dot{m}}  = \mu_o
\label{eq:muvar}
\eeq
Here, the conserved energy can be defined as with $\Eo$ as
$\rho_o=\frac{\partial \dot{m}}{\partial \rho}\rho$
and as with Equation (\ref{eq:importantrelation}) we have
\beq
\rho\beta=\beta_o\rho_o \s .
\eeq

Finally, we can find the variation in the remaining ``intensive'' quantity - the
pressure $p$ for systems which have particle flow.  
In order to have the system remain in mechanical equilibrium, the
pressure will have to vary throughout a substance in order to keep the
substance from flowing.  This requirement gives
\beq
\frac{dp}{dx}=-F(x)
\label{eq:pvar}
\eeq
where $F(x)$ is the force due to the interaction.  If the total $m$
is simply some potential, then one would have $F(x)=dm/dx$.

%
%

\section{gravitating perfect fluid}
\label{ap:scaling}
In this section, we will calculate the entropy of a spherically symmetric, self-gravitating
perfect fluid.  The field equations which govern the gravitating perfect fluid are well
known \cite{ov}.
Spherical symmetry implies that the metric takes the familiar form
\beq
ds^2=-e^{2\Phi}dt^2 + e^{2\Lambda} dr^2 + r^2 d\Omega^2
\eeq
where $\Phi$ and $\Lambda$ are functions of $r$.
The stress-energy tensor of the perfect fluid is given in terms of the energy density $\rho(r)$ and
radial pressure $p(r)$ by
\beq
T^{\mu\nu}=(\rho+p)u^\mu(r) u^\nu(r) + p(r) g^{\mu\nu}
\eeq
where $u^\mu(r)$ is the 4-velocity of the fluid and $g^{\mu\nu}$ is the metric.
Einstein's equation yield
\beq
e^{-2\Lambda}=1-2m(r)/r
\eeq
where
\beq
m(r)=\int_0^r 4\pi r^2 \rho dr
\eeq
and
\beq
\frac{d\Phi}{dr}=
\frac{m+4\pi r^3 p}{r(r-2m)}
\eeq
Outside the boundary of the fluid $r=R$, the functions $\Lambda$ and $\Phi$ reduce to
\beqa
e^{-2\Lambda(R)}
&=&
e^{2\Phi(R)} \nonumber\\
&=&
1-2 M/R
\eeqa
where $M\equiv m(R)$.


There are three conditions for equilibrium.  If the system is in
thermal equilibrium, it must obey the Tolman relation \cite{tolman}
\beq
T(r)=T_o e^{-\Phi(r)}
\eeq
where $T_o$ is the temperature as measured at infinity.
Likewise the chemical potential at any two points can be related by the redshift
to the value of the chemical potential on the boundary by 
\beq
\mu(r)=\frac{\mu(R)e^{\Phi(R)}}{e^{\Phi(r)}}
\eeq
The condition for hydorstatic equilibrium (i.e., no radial infalling of any fluid element) can be found
from local energy-momentum conservation
\beq
T^{\mu\nu}_{;\nu}=0
\eeq
which implies
\beq
(\rho + p)\phi,r=-p,r
\eeq
For a perfect fluid of constant density, this
leads to the well known Oppenheimer-Volkoff equation
\beq
\frac{dp}{dr}=-\frac{(\rho+p)(m+4\pi r^3 p)}{r(r-2m)}
\eeq
and simply balances the pressure gradiant with the force due to gravity until equilibrium is reached.

We can now calculate the total entropy of the system.  Since $s$ is the local entropy as measured by
observers in the rest frame of the fluid, we can integrate over the sphere to obtain the total entropy
$S$.  The appropriate volume element for a shell of thickness $dr$ is $dV=4\pi r^2 e^\Lambda dr$ and so
\beqa
S
&=&
\int_0^R dV s(r) \nonumber\\
&=& \frac{4\pi}{T_o}\int_0^R r^2 e^{\Phi+\Lambda} \left[\rho+p-\mu_o
  n \right] dr
\label{eq:entropygen}
\eeqa
where we have used the Gibbs-Duhem and Tolman relations. 

In general, we cannot solve this expression explicitely, however, for a perfect fluid
with constant energy density $\rho(r)=\rho_o$, and constant number density $n(r)=n_o$
the expressions for the metric and pressure are
well know \cite{Schwarzschild}\cite{mtw}.
\begin{displaymath}
  m(r)=\left\{\begin{array}{ll}
      (4\pi/3)\rho_o r^3 \s & r<R\\
      M= (4\pi/3)\rho_o R^3 \s & r>R\end{array} \right.
\end{displaymath}

\begin{displaymath}
  N(r)=\left\{\begin{array}{ll}
      (4\pi/3) n_o r^3 \s & r<R\\
      N= (4\pi/3)n_o R^3 \s & r>R\end{array} \right.
\end{displaymath}
%
%
%
\beq
e^\Phi=\frac{3}{2}k(R)-\frac{1}{2}k(r) \s r<R
\eeq
\beq
p=\rho_o \frac{k(r)-k(R)}{3k(R)-k(r)} \s r<R
\eeq
where $k(r)\equiv\sqrt{1-2Mr^2/R^3}$.
It is worth observing that the pressure at $r=0$
diverges as $2M/R\rightarrow 8/9$ and as a result, this limits the size of our sphere of fluid.

We can now compute the entropy using equation (\ref{eq:entropygen}).
\beqa
S&=&\frac{4\pi}{T_o}\int_0^R r^2dr\left[ \rho_o\frac{k(R)}{k(r)}-\mu n_o \frac{3k(R)-k(r)}{2k(r)} \right]
\nonumber\\
&=& \frac{3k(R)R}{4 T_o}
(1-\frac{N\mu{R}}{M})
\left[\sqrt{\frac{R}{2M}}\arcsin{\sqrt{\frac{2M}{R}}}-k(R)\right]
\eeqa

We see therefore, that the entropy is no longer an extensive quantity and does not scale linearly
with $N$ and $M$, as it would for a system whose entropy scales like the volume of the system.
However, if we expand our solution in terms of the gravitational coupling, $M/R$, then we find, to
0th order in $M/R$
\beq
S\equiv (M-\mu(R)N)/T_o
\eeq
and we recover the extensive scaling of the entropy.  To first order in $M/R$
\beq
S\equiv (M-\mu(R)N)(1-\frac{2}{5}\frac{M}{R})/T_p \s .
\eeq

One would like to calculate the entropy for various equations of state.
Unfortunately, one finds that for any realistic equation of state, the
system is of infinite size.  One can remedy this by having different
equations of state at different radii, but this makes calculating
the scaling behavior of the entropy completely meaningless, since
it would depend more on how one changed the equations of state
rather than on any properties of the states themselves.  
Another way of obtaining convergence, is to use the so-called
``polytropic'' equations of state such as
\beq
\rho=(\beta_o\mu_o+b)p^{\frac{a}{a+1}}
\eeq
However, they are not true
equations of state, and come from assuming that the matter is 
adiabatic as a function of $r$.  They are therefore not suitable
for states in thermal equilibrium.

\acknowledgments I would like to thank Jacob Bekenstein for
many helpful and interesting discussions on the subjects presented here.
It is also a pleasure to thank Michal Horodecki, Eliot Leib, and Don Page   
for their valuable comments.  J.O. acknowledges the support of
the Lady Davis Fellowship Trust, and ISF grant 129/00-1.  Much of this work
was originally presented at the 9th Canadian Conference on General Relativity
and Astrophysics, Edmonton, May, 2001.


\begin{thebibliography}{25}
\bibitem[{not()}]{notation}
\bibinfo{note}{Here $m$ is the total energy, and should not be confused with
  the magnetization -- the notation used will in general be more familiar to
  general relativists and may therefore be initially confusing to other
  theorists.}
\end{thebibliography}

\end{document}